**Molecular Beam Epitaxy Growth**
**of Transition Metal Dichalcogenide (Mo,Mn)$Se_2$ on 2D, 3D and polycrystalline substrates**

J. Kucharek,[1,*] R. Bożek,[1] W. Pacuski[1]

1. Institute of Experimental Physics, Faculty of Physics, University of Warsaw, Pasteura 5, 02-093 Warsaw, Poland

*corresponding author email: Julia.Kucharek@fuw.edu.pl

Abstract: Magnetic doping of 2D materials such as Transition Metal Dichalcogenides is promising for enhancement of magneto-optical properties, as it was previously observed for 3D diluted magnetic semiconductors. To maximize the effect of magnetic ions, they should be incorporated into the crystal lattice of 2D material rather than form separated precipitates. This work shows study on incorporating magnetic manganese ions into the $MoSe_2$ monolayers using molecular beam epitaxy. We test growth on various substrates with very different properties: polycrystalline $SiO_2$ on Si, exfoliated 2D hexagonal Boron Nitride flakes (placed on $SiO_2$/Si), monocrystalline sapphire, and exfoliated graphite (on tantalum foil). Although atomic force microscopy images indicate presence of MnSe precipitates, but at the same time, various techniques reveal effects related to alloying $MoSe_2$ with Mn: Raman scattering and photoluminescence measurements shows energy shift related to presence of Mn, scanning transmission microscopy shows Mn induced partial transformation of 1H to 1T' phase. Above effects evidence partial incorporation of Mn into $MoSe_2$ layer.

Keywords: $MoSe_2$, magnetism, manganese, hBN, photoluminescence, MBE, doping, alloy

## 1. Introduction

Electronic properties of two-dimensional (2D) materials associated with the additional degree of freedom originating from magnetic dopant spin orientation are being a subject of intense investigation due to their promising applications in 2D spintronic devices [1,2]. Mn-doped materials from Transition Metal Dichalcogenides (TMDs) family can possibly exhibit enhanced magnetooptical properties and fascinating magnetic phenomena as e.g. carrier mediated ferromagnetism [3,4,5,6]. However, the first technical obstacle to overcome is low solubility of Mn dopant. Limitations related to low solubility are observed in case of most of diluted magnetic semiconductors (except maybe for CdTe [7]), but are particularly visible in case of monolayer materials, e.g. $MoSe_2$ on graphene, where Mn tends to form MnSe clusters [8]. Solubility limit is slightly less visible in case of 2D materials in form of multilayers ($MoSe_2$ on $SiO_2$ [9]) or nanostructures ($MoS_2$ [10]). Additionally, reactive substrates can bind manganese atoms and therefore block incorporation into dichalcogenide monolayer [11]. However, best optical properties are reserved for TMDs in form of monolayers, which are grown on hBN [12,13] or encapsulated in hBN [14,15,16]. Consequently, in our research on growth of $MoSe_2$ with manganese we focus on synthesis of monolayers. Moreover to a set of tested substrates ($SiO_2$, graphite and sapphire) we added exfoliated hBN.

## 2. Samples

Studied $(Mo,Mn)Se_2$ layers were grown using molecular beam epitaxy in growth chamber dedicated to II-VI semiconductors. Two kinds of molecular sources were used: effusion cells for selenium and manganese, where the flux of element was adjusted by changing temperature of cell, and electron-beam source with a rod, where the molybdenum flux was adjusted by changing power of the source.

In the same conditions were obtained molybdenum diselenide ($MoSe_2$) – reference sample and a series of manganese-molybdenum diselenide ($(Mo,Mn)Se_2$) samples. Nominal concentration of manganese was estimated based on the amount of deposited Mo and Mn.

The following substrates were used: Si with polycrystalline $SiO_2$ buffer ($SiO_2$/Si), exfoliated hexagonal Boron Nitride flakes deposited on $SiO_2$/Si wafer (hBN/$SiO_2$/Si), $Al_2O_3$ (sapphire) and graphite on tantalum foil. Each growth process was characterised by duration, molecular fluxes, temperature of substrate during growth and annealing. Before growth the substrate was degas at temperature equal to predicted post-growth annealing temperature. Details of growth are gathered in Tables 1 and 2.

| Manganese concentration | | Substrate | | | |
|---|---|---|---|---|---|
| nominal Mn content [%] | Mn source temperature [°C] | $SiO_2$/Si | Sapphire | hBN/$SiO_2$/Si | Graphite |
| 0 | 0 | UW1396 | UW1252 | UW1668 | UW1536 |
| 7.5 | 590 | UW1663 | | UW1663 | UW1535 |
| 12.4 | 600 | | UW1267 | | |
| 28.9 | 625 | | UW1266 | | |
| 52.5 | 650 | UW1401 | UW1264 | | |
| 74.0 | 675 | UW1400 | UW1260 | | |
| 87.4 | 700 | UW1399 | UW1258 | | |
| 94.2 | 725 | UW1398 | UW1257 | | |
| 97.3 | 750 | UW1397 | UW1256 | | |
| 98.7 | 775 | | UW1255 | | |
| 99.4 | 800 | | UW1254 | | |
| 100.0 | 850;700;590 | UW1040 | UW1203 | UW1662 | |

Table 1. Table of studied samples assigned to the substrates and the concentration of manganese. Each sample id is in the form UWXXXX.

| Sample ID | Duration of growth [min] | Temperature of sub. during growth [°C] | Annealing of sample after growth [°C] | Molecular sources |
|---|---|---|---|---|
| UW1040 | 45 | 300 | non | Mn@850 °C, Se@220 °C |
| UW1203 | 42 | 450 | non | Mn@700 °C, Se@220 °C |
| UW1252-UW1267 | 20 | 300 | 470 | Mo@160 W, Se@220 °C |
| UW1396-UW1401 | 50 | 300 | 510 | Mo@160 W, Se@220 °C |
| UW1668 | 300 | 300 | 700 | Mo@115 W, Se@220 °C |
| UW1535,UW1536 and UW1663 | 180 (40 min with Mn) | 400 | 700 | Mo@130 W, Se@220 °C |
| UW1662 | 40 | 300 | non | Mn@590 °C, Se@220 °C |

Table 2. Growth parameters of studied samples: time of growth, substrate temperature during growth, annealing temperature and parameters of molecular sources. Parameters for Mo and Se sources and time of growth were in each case optimized for 1 monolayer (except samples: UW1397-UW1401), it may slightly vary due to recalibration.

## 3. Results and discussion

### 3.1 Growth on polycrystalline $SiO_2$/Si and crystalline sapphire

AFM images of a representative (Mo,Mn)$Se_2$ sample with nominal concentration of Mn equal 7.5% is presented in the Fig. 1. Next to it, there are presented AFM images of two reference samples: pure MnSe and pure $MoSe_2$. When comparing these tree sets of pictures, we observe two features: relatively rough surface and high point aggregates. Rough surface is observed for each sample and we attribute it to roughness of $SiO_2$ covered eventually by very thin (1 ML) $MoSe_2$ or (Mo,Mn)$Se_2$ layers that reflects substrate roughness. Since high point aggregates are observed only for samples containing Mn, and they are similar to MnSe known from literature [9,17], we attribute them to MnSe nanocrystals, and we conclude that we observe phase segregation. What is also worth mentioning, despite relatively high point aggregates, the average roughness (AR) of samples containing manganese is lower than the ones without it. Taking into consideration images in Fig. 1. the change can be up to around 20%. This effect can be associated with the surfactant properties of manganese that can effectively decrease crystallite size even 4 times in many various materials [18-20].

For photoluminescence study we have prepared a series of samples on $SiO_2$/Si substrate with various amount of deposited Mn while amounts of Mo and Se were kept constant. Amount of $MoSe_2$ was chosen to be under 1 monolayer, and then monolayer was fulfilled using various amounts of Mn. Thickness of $SiO_2$ was 90 nm to optimize PL studies due to positive interferences.

Fig. 2a. presents room temperature photoluminescence study of $MoSe_2$ and (Mo,Mn)$Se_2$, excited by 532 nm laser. The addition of manganese has not quenched the intensity of PL related to $MoSe_2$, on the contrary Mn-infused samples exhibit slightly higher PL intensity than pure ones. This can be explained in a following way: the experiment started with growing an incomplete $MoSe_2$ monolayer resulting in weaker PL, and then, in subsequent samples, increasing amount of manganese has led to complementing the monolayer, which improved the intensity of the photoluminescence. The second important result related to PL is the shifting of the (Mo,Mn)$Se_2$ peaks relative to the $MoSe_2$ peak. This means that manganese influences the band gap energy in the obtained material, which suggests that Mn is incorporated into the crystal structure, at least partially. Additionally, Fig. 2b also shows in different scale that PL of MnSe is at higher energy than $MoSe_2$ or $(Mo,Mn)Se_2$. The origin of the observed PL is divers in the case of various materials. In the case of $MoSe_2$ and $(Mo,Mn)Se_2$, PL close to 1.55 - 1.6 eV is related to excitonic transitions and semiconductor energy band gap. Very differently, the energy gap of MnSe is in UV range (see a drop of transmittance of MnSe close to 3 eV in Fig. 3), and PL observed at 1.8 eV is related to d-d transitions of Mn hybridized with $4p^4$ states of Se [17,21,22].

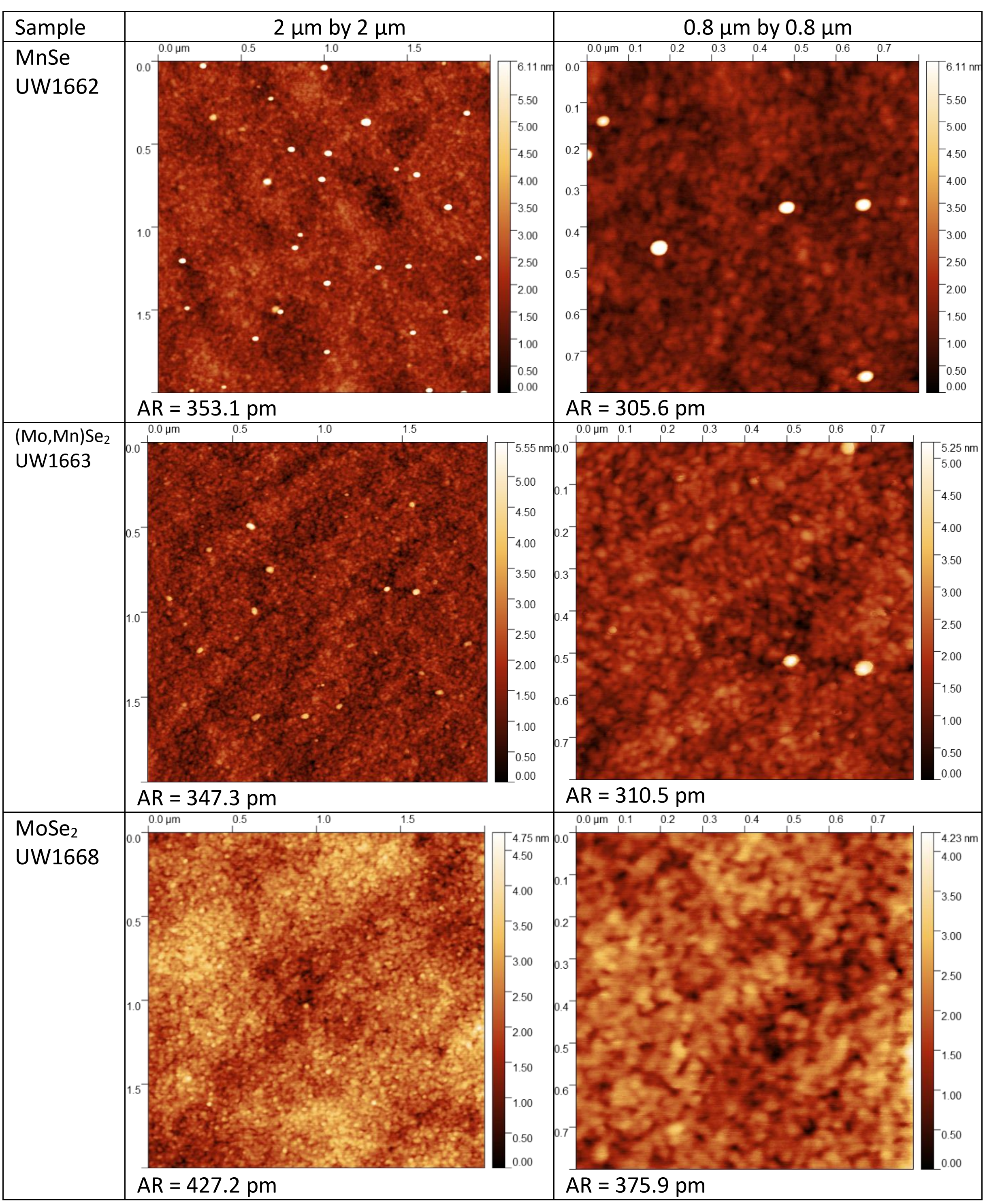

Figure 1. Comparison of AFM images of MnSe, $(Mo,Mn)Se_2$ and $MoSe_2$ layers on $SiO_2$ substrate in two different magnifications – left 2 µm by 2 µm, right 800 nm by 800 nm. Under every image the average roughness was noted as AR=XX.

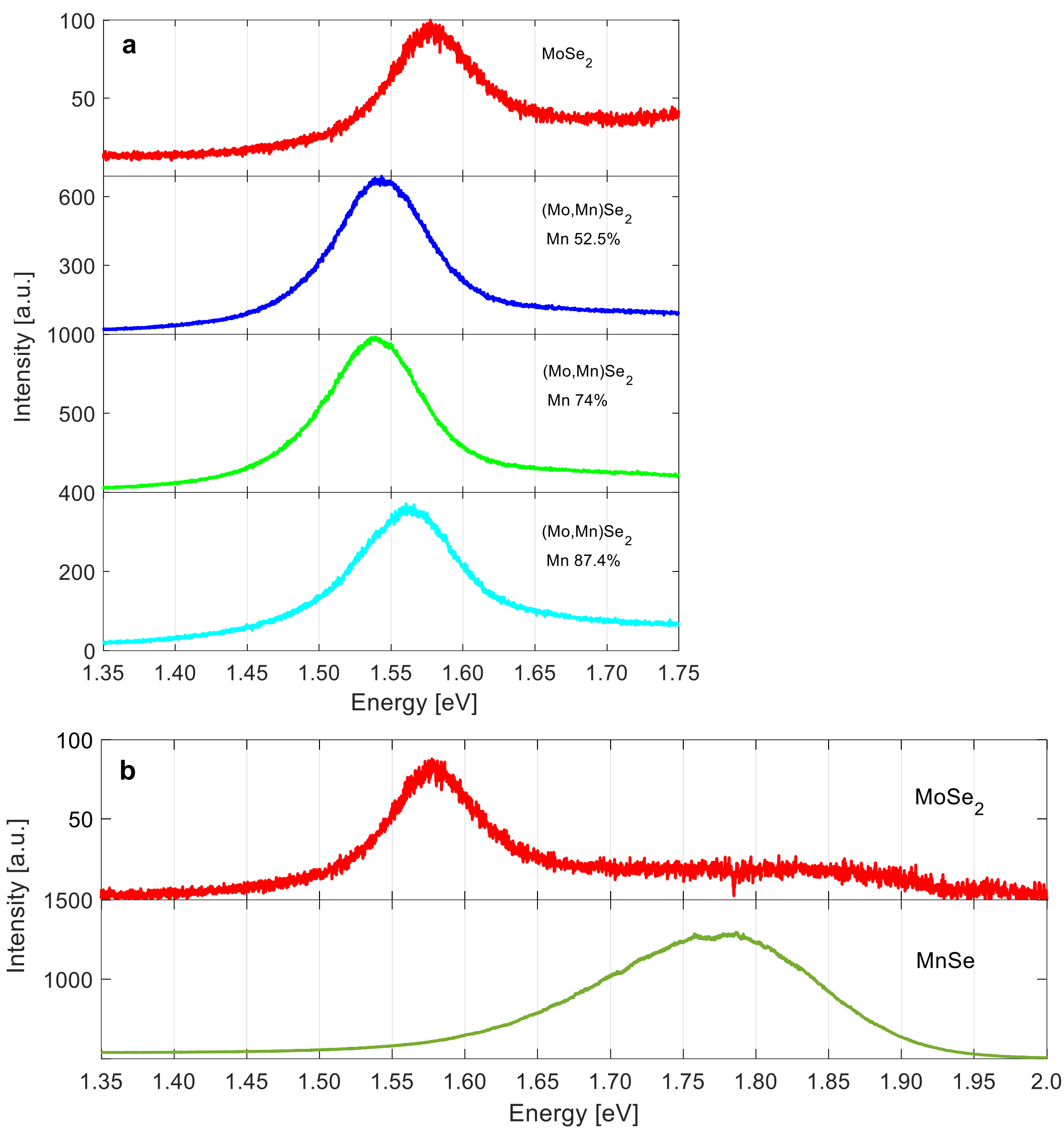

Figure 2. a - Room temperature PL spectra of $MoSe_2$ and $(Mo,Mn)Se_2$ samples grown on $SiO_2$/Si substrate. Red, upper line represents $MoSe_2$ spectrum; tree lower lines show spectra for $(Mo,Mn)Se_2$ selected samples; b - comparison of room temperature PL spectra of MnSe and $MoSe_2$ samples grown on $SiO_2$/Si substrate, in wider spectral range. A green laser emitting a wavelength of 532 nm was used for excitation. UW1396 UW1401, UW1400, UW1399 and UW1040 samples were measured.

Effects related to Mn incorporation to the lattice can be also observed in transmittance measurements, in particular for samples grown on transparent substrate such as sapphire. Results are presented in Fig. 3 and show that from transmittance we can learn not only how exciton A shifts, but we can also observe exciton B. Similarly as in the discussion of PL spectra, the excitonic redshift occurs, suggesting that manganese influences the band gap energy in the obtained material. Moreover, we see that only in the situation of almost pure MnSe, in sample with over 94% of nominal concentration of Mn, $MoSe_2$ features related to $MoSe_2$ excitons are quenched. This can be understood if we assume partial incorporation of Mn in $(Mo,Mn)Se_2$ and substantial precipitation of Mn in form of MnSe.

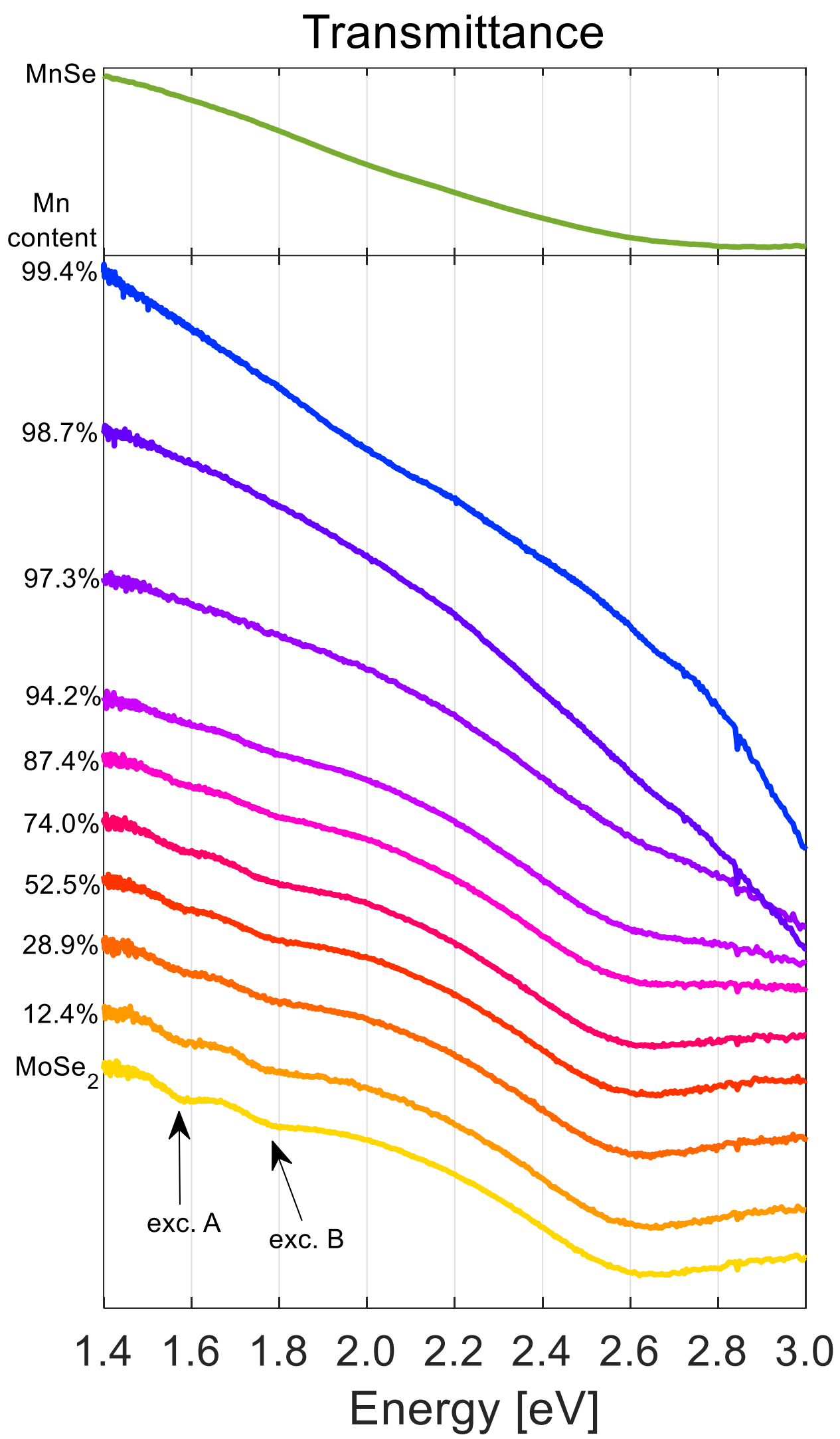

Figure 3. Transmittance spectra of: top panel – MnSe and bottom panel - $MoSe_2$ and 8 $(Mo,Mn)Se_2$ samples grown on sapphire. Yellow, bottom line shows $MoSe_2$ reference sample spectrum, the lines above are spectra from samples with increasing amount of manganese, ordered from the lowest concentration, beginning from $MoSe_2$, the top line is transmittance spectra from MnSe reference sample. Spectra are shifted for clarity. Measurements were performed in room temperature using white light source. UW1203, UW1267, UW1266, UW1264, UW1260, UW1258, UW1257, UW1256, UW1255, UW1254 and UW1252 samples were employed.

To study effect of Mn doping, we engaged also room temperature Raman scattering excited by 532 nm laser. We observed that the characteristic Raman line of $MoSe_2$ at 241 $cm^{-1}$ evolves with increasing amount of Mn, namely it shifts towards low energies, as shown in Fig. 4 for samples grown on sapphire. Similar effect has been already reported in Ref. [9] for $(Mo,Mn)Se_2$ grown on $SiO_2$ and interpreted as a consequence of Mn-related strain or Mn intercalation in multilayer structure. In each situation, presence of Mn very close to $MoSe_2$ lattice is confirmed. We interpret nonmonotonic dependence of line energy versus nominal concentration of Mn, as shown in Fig. 4b, as being a consequence of Mn solubility limit. Despite $(Mo,Mn)Se_2$ Raman lines are observed at lower energy than $MoSe_2$, line related to MnSe is observed at higher energy than $MoSe_2$ therefore there is no continues change from one material to another. Importantly, MnSe line exhibit much weaker intensity than $MoSe_2$ line, what suggest that peaks in $(Mo,Mn)Se_2$ originate from $MoSe_2$ lattice, even in samples with relatively high Mn concertation.

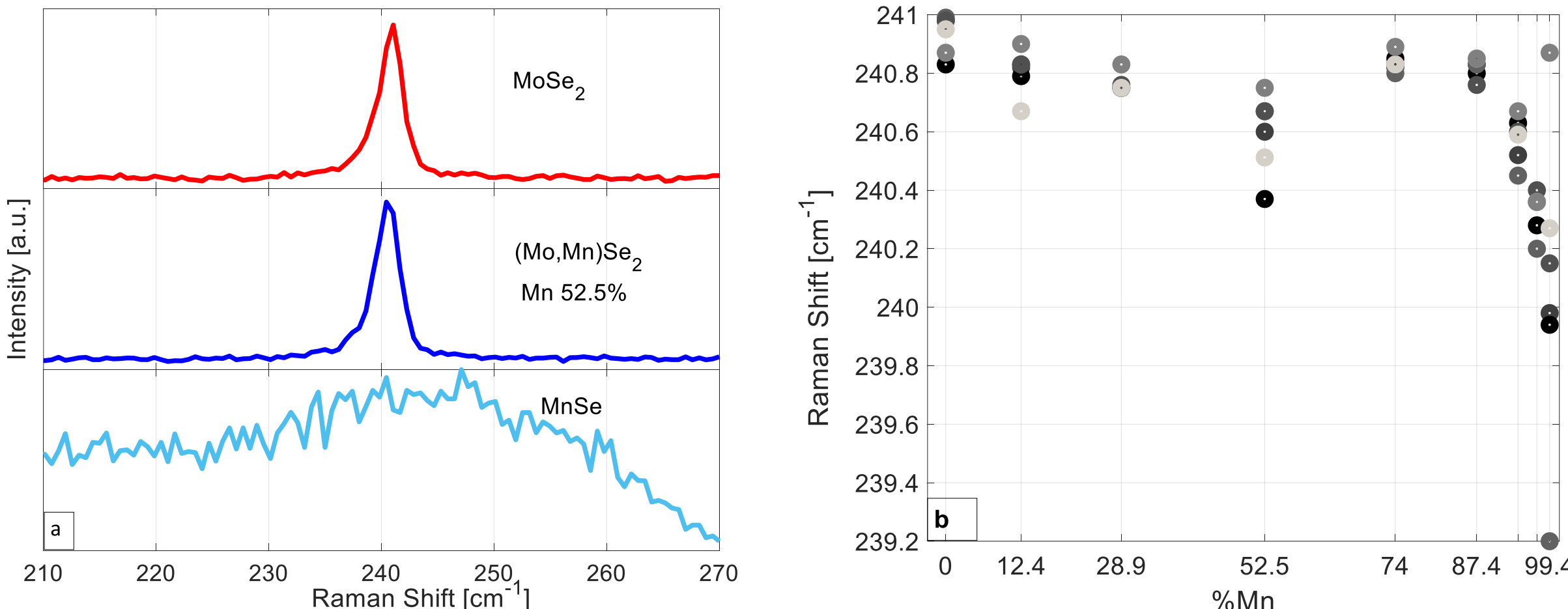

Figure 4. Raman scattering room temperature spectra of $MoSe_2$, $(Mo,Mn)Se_2$, and MnSe samples. a – red, upper line represents $MoSe_2$ $A_{1g}$ vibrational mode; dark blue, middle line shows the same mode for $(Mo,Mn)Se_2$, with nominal 52.5 % Mn ions content, that is slightly shifted towards lower energies, and light blue, bottom line shows wide pick from MnSe, that potentially gathers many different modes. b – $A_{1g}$ vibrational mode energies in respect to Mn concentration in subsequent samples: $MoSe_2$ and 8 various $(Mo,Mn)Se_2$ samples. For each sample, measurement has been done for a few spots on a surface, represented by various experimental points. A green laser emitting a wavelength of 532 nm was used for excitation. UW1252, UW1267, UW1266, UW1264, UW1260, UW1258, UW1257, UW1256, UW1255, UW1254, UW1203 samples were employed.

### 3.2 Growth on $hBN/SiO_2/Si$

Exfoliated hBN is the inalienably perfect substrate for growing $MoSe_2$ with good optical properties [12], but in our case it also exhibit the advantage of almost perfect flatness facilitating AFM imaging of grown flakes, as shown in Fig. 5. One can find small 1 ML thick flakes of TMD and a few nm high points similar to observed already on $SiO_2$ and interpreted as MnSe. Therefore the phase segregation is directly observed. When it comes to surface roughness, one can see that in images in Fig. 5. the main roles play very smooth area of uncovered hBN substrate and flat $MoSe_2$ flakes area. These two elements respectively, make MnSe and $MoSe_2$ surfaces less rough than their equivalents in Fig. 1. Analizing $(Mo,Mn)Se_2$ surface on hBN, one can perceive that it is composed of plenty of slight $MoSe_2$ flakes (size below 50 nm) and relatively high MnSe nanocrystals (diameter below 20 nm). All that generate increased surface roughness in comparison to its counterpart from Fig. 1.

To check if flat TMD flakes contain Mn, we performed PL measurements at helium temperatures, as show in Fig. 6. Also this time, the excitonic PL peak is slightly shifted under the influence of manganese ions, suggesting at least partial incorporation of Mn into $MoSe_2$ lattice. Comparing to very sharp and narrow lines of $MoSe_2$ grown on hBN, $(Mo,Mn)Se_2$ grown on hBN exhibits much wider lines, moreover charged and neutral exciton lines are not resolved anymore.

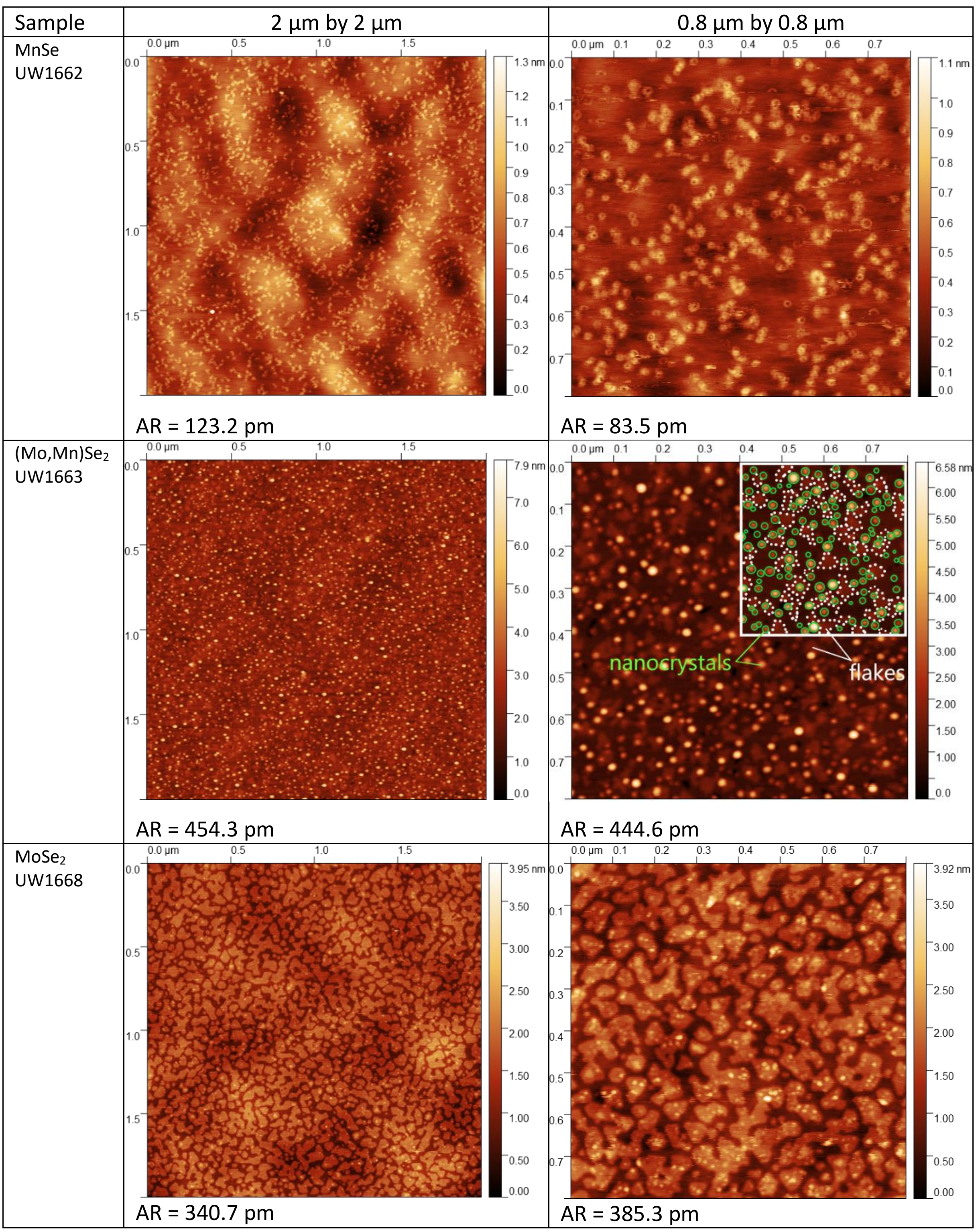

Figure 5. Comparison of AFM images of MnSe, (Mo,Mn)$Se_2$ and $MoSe_2$ layers on hBN substrate. Under every image the average roughness was noted as AR=XX.

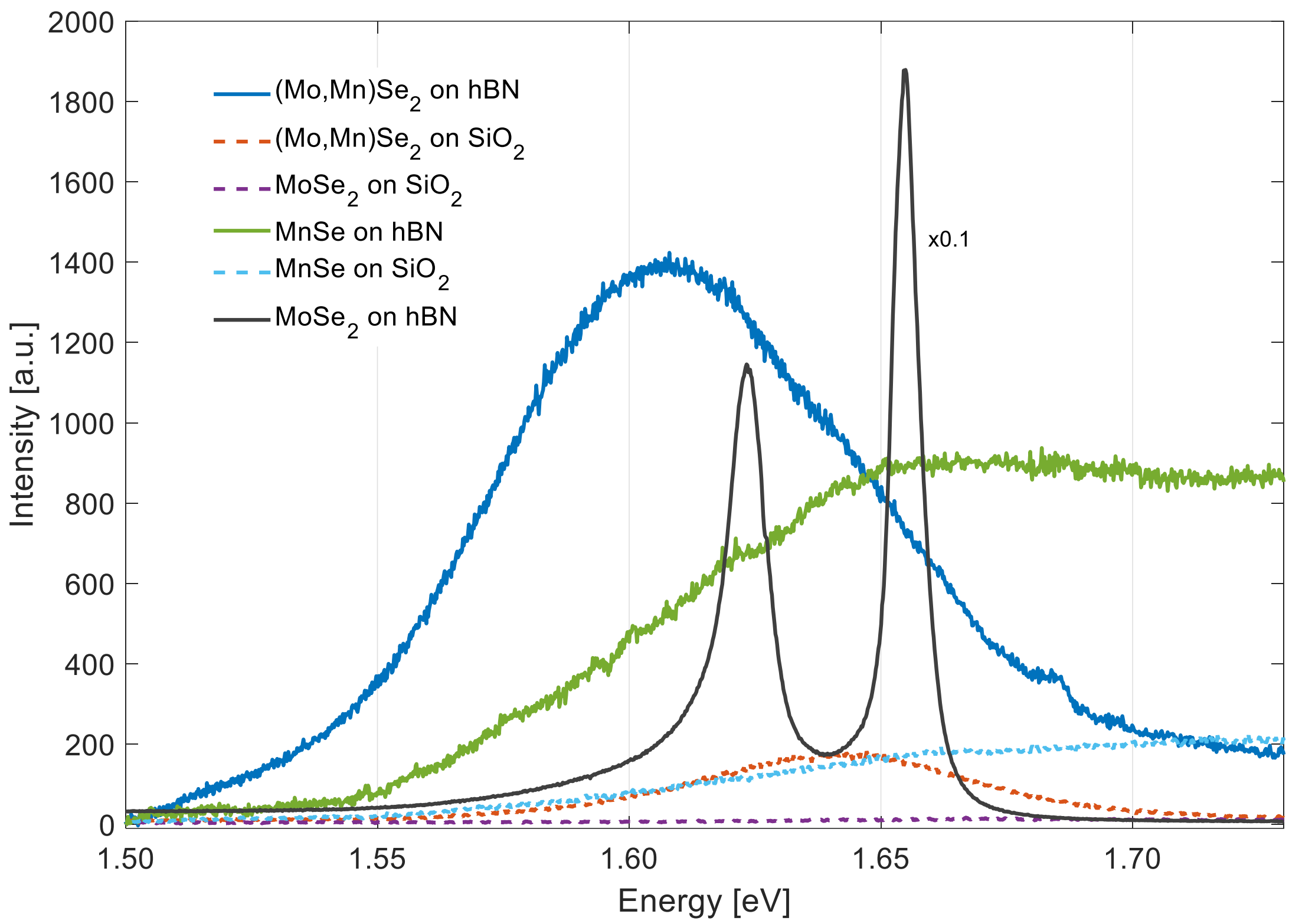

Figure 6. Photoluminescence spectra of the (Mo,Mn)$Se_2$ sample: blue – on hBN and red on $SiO_2$; PL spectra of reference MBE samples: $MoSe_2$ (black and violet) and MnSe (green and cyan) grown on hBN and $SiO_2$. The measurements were performed at 10 K. A green laser emitting a wavelength of 532 nm was used for excitation. UW1663, UW1668 and UW1662 were used for measurements.

### 3.3 Growth on Graphite

For better understanding of atoms distribution in studied compound, we have grown $MoSe_2$ and (Mo,Mn)$Se_2$ monolayers on exfoliated graphite - ultra-flat, conductive substrate enabling Scanning Transmission Microscopy (STM) investigations. Results of STM imaging of (Mo,Mn)$Se_2$ are shown in Fig. 7b. and 8. They reveal coexistence of two crystallographic TMD phases: 1H – the most popular and stable semiconductor one, with trifold symmetry and 1T' – more rare, semimetallic phase, with twofold symmetry. Importantly, STM images didn't reveal similar 1T' phase for pure $MoSe_2$ without Mn. Pure $MoSe_2$ sample is shown in Fig. 7a, where STM image is dominated by triangular inversion domains [23].

1T' phase for $MoSe_2$ is a very rare case in a literature, observed only in very particular conditions. The article by Feng Cheng et al. [24] showed how this phase is induced by depositing an additional molybdenum layer on a gold substrate with selenium reconstitution. Molybdenum creates elevated platforms and then intercalates between selenium and gold atoms. The molybdenum and selenium atoms are used up to form $MoSe_2$. The same report also explains the phenomenon by calculations showing that the 1T' phase is favored when the growth takes place on an additional layer of molybdenum, which is a transition metal. This makes it 0.851 eV / $MoSe_2$ (0.851 eV per one $MoSe_2$ molecule) more stable than 1H-$MoSe_2$. By means of the Bader charge analysis it was shown that the charge transfer between $MoSe_2$ and the d Mo shell stabilizes the 1T' phase and allows it to be grown as dominant. On this basis, we interpreted that in the case of the studied sample (Mo,Mn)$Se_2$, the role of the additional layer of Mo atoms was taken over by the added Mn, which also belongs to the transition metals and has a sequence of electron shells similar to Mo. Which caused the coupling between the d-manganese coating and the growing material, and as a result, the formation of the 1T' phase.

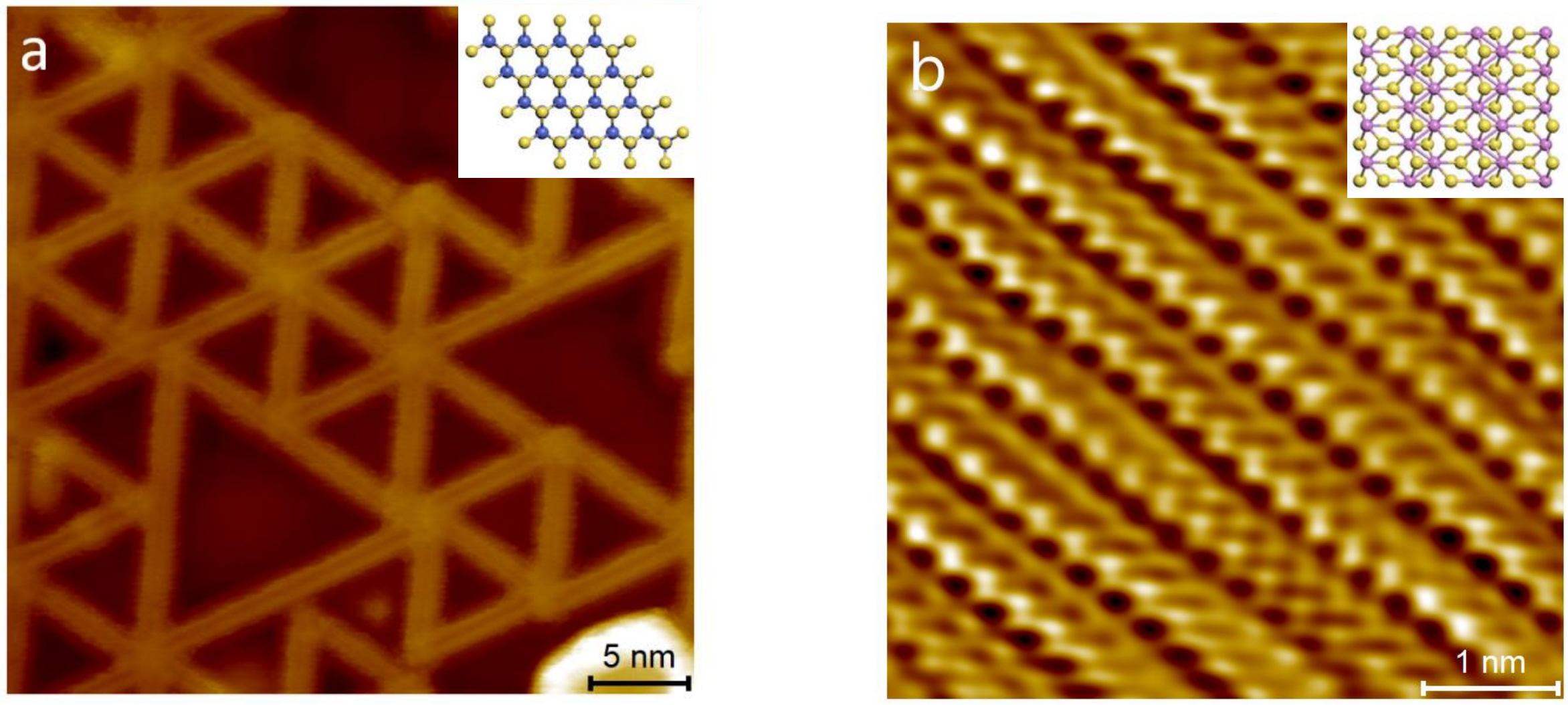

Figure 7. STM image of the surface of samples grown on graphite: a - 1H phase of $MoSe_2$ (UW1536 sample) and b – 1T' phase of $(Mo,Mn)Se_2$ (UW1535 sample). Insets with atomic structures come from Y. Sim et al. [23].

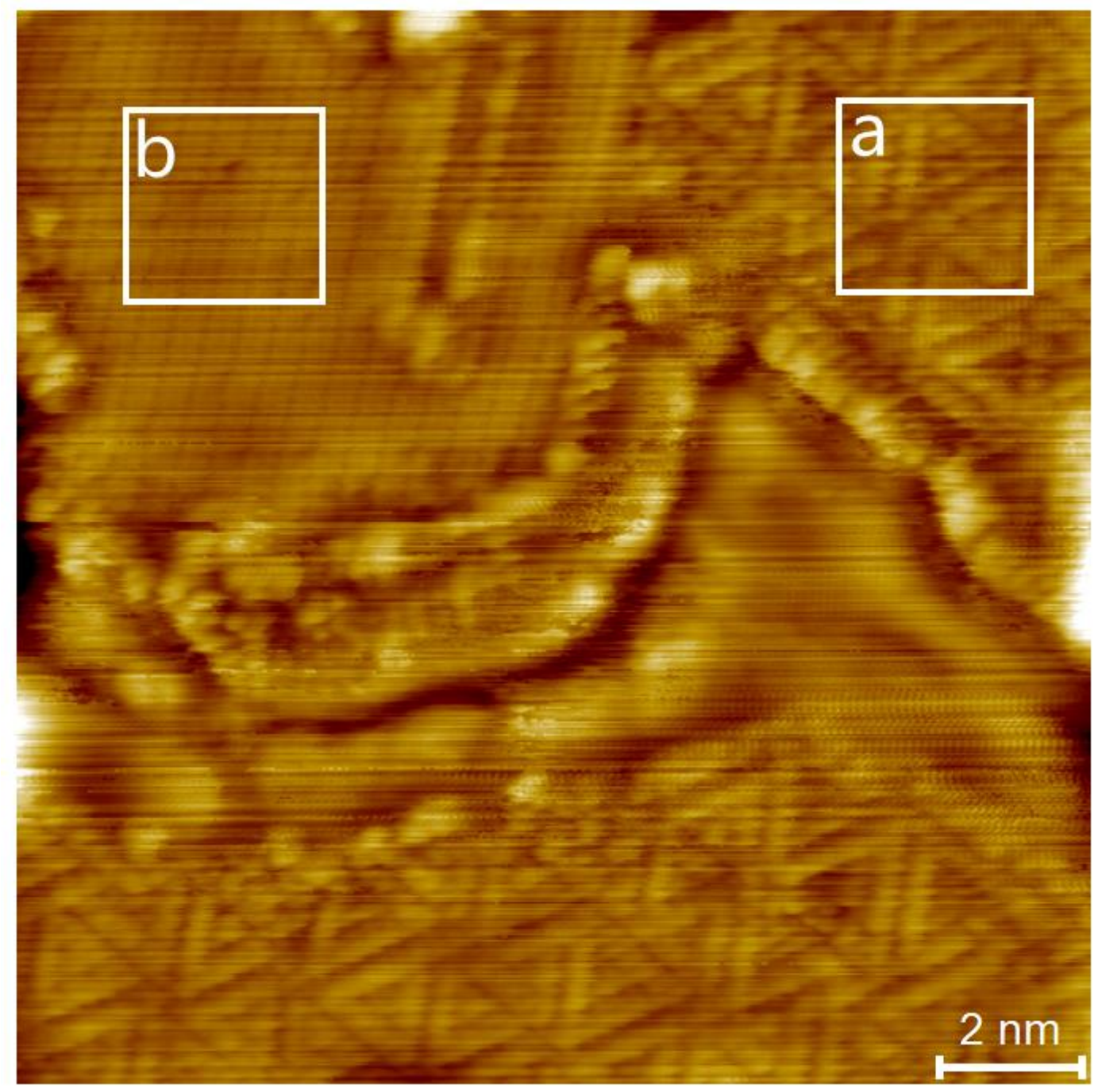

Figure 8. STM image of $(Mo,Mn)Se_2$ surface (UW1535 sample). There are strip domains (left frame) coexisting with triangular domains (right frame), both corresponding to different crystallographic phases. Triangular domains are due to typical defects [24] in the most stable form - 1H - while the striped area corresponds to the less stable 1T' phase [25].

## 4. Conclusions

Presented work shows the growth of molybdenum diselenide alloyed with manganese magnetic ions on various substrates including polycrystalline $SiO_2$, crystalline 3D sapphire and 2D substrates hBN and graphite. For each substrate we observe impact of Mn on properties of $MoSe_2$. Optical spectroscopy (photoluminescence, transmittance and Raman scattering) evidences that Mn induces shifts and broadening of spectral lines and consequently reveals at least partial incorporation of Mn into $MoSe_2$ lattice. In particular we observe lowering of energy gap of $(Mo,Mn)Se_2$ with increasing concentration of Mn. At the same time, AFM imaging reveals also at least partial segregation of MnSe nanocrystals from 2D $MoSe_2$. AFM pictures provides also an example of surface flattening under the influence of Mn addition. The change in average roughness can reach around 20% on relatively rough substrates such as $SiO_2$. In case of perfectly flat hBN substrate, addition of Mn only increases roughness. STM imaging reveals that Mn addition leads to the formation of two kind of domains with typical 1H crystallographic phase and rather rare 1T' crystallographic phase, what can be promising in future investigations of electron transport in $MoSe_2$ e.g. in terms of Weyl semimetals [26].

## Acknowledgments

We acknowledge the financial support from National Science Centre Poland projects no. 2017/27/B/ST5/02284 and 2021/41/B/ST3/04183. This work was also supported by the Polish Ministry of Science and Higher Education as research grant "Diamentowy Grant" under decision DI2019 006149.